\documentstyle[aps,prb,multicol,epsf]{revtex}

\begin{document}
\draft

\title{A-type Antiferromagnetic and C-type Orbital-Ordered State 
in LaMnO$_3$ \\ Using Cooperative Jahn-Teller Phonons}

\author{Takashi Hotta, Seiji Yunoki, Matthias Mayr, and Elbio Dagotto}

\address{National High Magnetic Field Laboratory, 
Florida State University, Tallahassee, Florida 32306}

\date{\today}

\maketitle

\begin{abstract}
The effect of Jahn-Teller phonons on the magnetic and orbital
structure of LaMnO$_3$ is 
investigated using a combination of relaxation and Monte Carlo techniques 
on three-dimensional clusters of MnO$_6$ octahedra.
In the physically relevant region of parameter space for LaMnO$_3$,
and after including small corrections due to tilting effects,
the A-type antiferromagnetic 
and C-type orbital structures were stabilized, in agreement 
with experiments. 
\end{abstract}


\begin{multicols}{2}
\narrowtext

The theoretical understanding 
of manganese oxides is among the most challenging current areas of
research in condensed matter physics.
\cite{review} Experimental studies of manganites have revealed
a rich phase diagram originating from the competition between
charge, spin, and orbital degrees of freedoms.
A simple starting framework for Mn-oxide investigations is 
contained in the double-exchange (DE) ideas, where ferromagnetism
induced by hole doping arises from the
optimization of the hole kinetic energy.\cite{DEmodel} In addition,
recent results using the one-orbital model 
revealed a more complicated ground state,
with phase separation tendencies strongly competing with 
ferromagnetism,\cite{Yunoki} leading to a potential
explanation of the Colossal Magneto-Resistance effect.\cite{Moreo}

However, to understand the fine details of the phase diagram of manganites
the one-orbital model is not sufficient since the highly nontrivial
A-type spin 
antiferro (AF) and C-type orbital structures observed experimentally
in the undoped material 
LaMnO$_3$\cite{LaMnO3} cannot be properly addressed in such a simple context.
Certainly two-orbital models are needed to consider the nontrivial state
of undoped manganites. In this framework 
the two-band model 
without phonons has been studied before, and the
importance of the strong Coulomb repulsion has been remarked for the
appearance of the A-AF state.
\cite{Coulomb,Mizokawa,Ishihara,Maezono}
However,
Coulombic approaches have presented conflicting results regarding the
orbital order that coexists with the A-type spin state, with several
approaches predicting 
G-type orbital order, which is not observed in practice.
In addition, many experiments suggest that
Jahn-Teller (JT) phonons are important in manganites and, thus, at present 
it is unclear
whether the dominant interaction between electrons in Mn-oxides should be
considered Coulombic or phonon mediated. For this
reason it is important to analyze if a purely JT-phononic calculation is able 
to reproduce the experimental properties of undoped manganites, goal
that provides the  motivation for the present paper.
The main result observed in this effort
is that A-AF order, in combination with a C-type orbital arrangement,
is indeed induced by JT-phonons 
in realistic parameter regions for LaMnO$_3$, namely
large Hund-coupling between e$_{\rm g}$-electrons and t$_{\rm 2g}$-spins,
small AF interaction between t$_{\rm 2g}$-spins,
and strong electron-lattice coupling.
This shows that JT-based calculations can lead to correct
qualitative predictions for manganites, actually improving on
purely Coulombic approaches in the orbital sector. 

To carry out the calculations note that experiments have revealed an
orbital 
structure tightly related to the JT-distortion of the
MnO$_6$ octahedron. 
If each JT-distortion would occur independently, optimal orbitals can be 
determined by minimizing the kinetic and interaction energy
of the e$_{\rm g}$-electrons, as in models with only Coulomb
interactions. 
However, oxygens are shared between adjacent MnO$_6$ 
octahedra, indicating that the JT-distortions occurs {\it cooperatively}. 
Particularly in the undoped situation, all MnO$_6$ octahedra exhibit 
JT-distortions, indicating that such a  cooperative effect is 
important.\cite{Kanamori}
Thus, in order to understand the magnetic and orbital structures of LaMnO$_3$, 
the electron and lattice 
systems must be optimized simultaneously.
However, not much effort has been devoted to the microscopic treatment of 
the cooperative effect,\cite{Allen} although the JT-effect in the Mn-oxides
has been addressed by several groups.
\cite{JTpolaron,Koizumi,Yunoki2}
To remedy this situation, here a computational
investigation of cooperative JT-phonons in
manganites is carried out,
focusing on the $n=1$ density, where $n$ is the electron number per site.

The motion of e$_{\rm g}$-electrons tightly coupled to the 
localized t$_{\rm 2g}$-spins and the local distortions of the 
MnO$_6$ octahedra is described by
\begin{eqnarray}
  H &=& -\sum_{{\bf ia}\gamma \gamma' \sigma}
  t^{\bf a}_{\gamma \gamma'} c_{{\bf i} \gamma \sigma}^{\dag} 
  c_{{\bf i+a} \gamma' \sigma}   
  - J_{\rm H} \sum_{{\bf i}\gamma\sigma \sigma'}
  {\bf S}_{\bf i} \cdot c^{\dag}_{{\bf i} \gamma \sigma}
  \bbox{\sigma}_{\sigma \sigma'} c_{{\bf i} \gamma \sigma'} \nonumber \\
  &+&  \lambda \sum_{{\bf i} \sigma\gamma\gamma'}
  c_{{\bf i} \gamma \sigma}^{\dag}
  (Q_{1{\bf i}} \sigma_0 +Q_{2{\bf i}} \sigma_1 + Q_{3{\bf i}}\sigma_3)
  _{\gamma\gamma'} c_{{\bf i} \gamma' \sigma} \nonumber \\
  &+& J' \sum_{\langle {\bf i,j} \rangle}
  {\bf S}_{\bf i} \cdot {\bf S}_{\bf j} 
  +(1/2)\sum_{\bf i}(\beta Q_{1{\bf i}}^2
  +Q_{2{\bf i}}^2+Q_{3{\bf i}}^2),
\end{eqnarray}
where $c_{{\bf i}a \sigma}$ ($c_{{\bf i} b \sigma}$) is
the annihilation operator for an e$_{\rm g}$-electron with spin $\sigma$ 
in the $d_{x^2-y^2}$ ($d_{3z^2-r^2}$) orbital at site ${\bf i}$.
The vector connecting nearest-neighbor (NN) sites is ${\bf a}$, 
$t^{\bf a}_{\gamma \gamma'}$ is the hopping amplitude between $\gamma$- and 
$\gamma'$-orbitals connecting NN-sites along the ${\bf a}$-direction 
via the oxygen 2$p$-orbital, 
$J_{\rm H}$ is the Hund coupling,
${\bf S}_{\bf i}$ the localized classical t$_{\rm 2g}$-spin 
normalized to $|{\bf S}_{\bf i}|=1$,
and $\bbox{\sigma}=(\sigma_1, \sigma_2, \sigma_3)$ are the Pauli matrices.
The dimensionless electron-phonon coupling constant is $\lambda$,
$Q_{1{\bf i}}$ denotes the dimensionless distortion for the breathing mode
of the MnO$_6$ octahedron, 
$Q_{2{\bf i}}$ and $Q_{3{\bf i}}$ are, respectively, 
JT distortions for the $(x^2-y^2)$- and $(3z^2-r^2)$-type modes, 
and $\sigma_0$ is the unit matrix. 
$J'$ is the AF-coupling between NN t$_{\rm 2g}$-spins,
and $\beta$ is a parameter to be defined below.

To account for the cooperative nature of the JT-phonons,
the normal coordinates for distortions of the MnO$_6$ 
octahedron are written\cite{Allen} as
$Q_{1 {\bf i}}=(1/\sqrt{3})(L_{\bf xi}+ L_{\bf yi}+L_{\bf zi})$,
$Q_{2 {\bf i}}=(1/\sqrt{2})(L_{\bf xi}- L_{\bf yi})$,
and 
$Q_{3 {\bf i}}=(1/\sqrt{6})(2 L_{\bf zi}- L_{\bf xi}-L_{\bf yi})$,
where $L_{\bf ai}$ denotes the distance between
neighboring oxygens along the ${\bf a}$-direction, given by
$L_{\bf ai} = L_{\bf a} +
(u_{\bf i}^{\bf a}-u_{\bf i-a}^{\bf a})$.
Here, $L_{\bf a}$ is the length between Mn-ions 
along the ${\bf a}$-axis
and $u_{\bf i}^{\bf a}$ denotes the deviation of oxygen from the 
equilibrium position along the Mn-Mn bond in the 
${\bf a}$-direction.\cite{comment}
In general, $L_{\bf a}$ can be different for each direction,
depending on the bulk properties of the lattice.
Since the present work focuses on the microscopic mechanism for A-AF 
formation in LaMnO$_3$, 
the undistorted lattice with $L_{\bf x}=L_{\bf y}=L_{\bf z}$ 
is treated first, and then corrections will be added.
In the cubic undistorted lattice, the hopping amplitudes are given by
$t_{\rm aa}^{\bf x}=-\sqrt{3}t_{\rm ab}^{\bf x}
=-\sqrt{3}t_{\rm ba}^{\bf x}=3t_{\rm bb}^{\bf x}=t$ for 
the ${\bf x}$-direction,
$t_{\rm aa}^{\bf y}=\sqrt{3}t_{\rm ab}^{\bf y}=\sqrt{3}t_{\rm ba}^{\bf y}=
3t_{\rm bb}^{\bf y}=t$ for the ${\bf y}$-direction, 
and $t_{\rm bb}^{\bf z}=4t/3$ with 
$t_{\rm aa}^{\bf z}=t_{\rm ab}^{\bf z}=t_{\rm ba}^{\bf z}=0$ 
for the ${\bf z}$-direction.
The energy unit is $t$.
The parameter $\beta$ is defined as 
$\beta=(\omega_{\rm br}/\omega_{\rm JT})^2$,
where $\omega_{\rm br}$ and $\omega_{\rm JT}$ are the vibration energies 
for manganite breathing- and JT-modes, respectively, assuming that 
the reduced masses for those modes are equal. 
Using experimental results and band-calculation data for 
$\omega_{\rm br}$ and $\omega_{\rm JT}$,\cite{Iliev}
it can be shown that $\beta \approx 2$. However, the results presented here 
are basically unchanged as long as $\beta$ is larger than unity.

To study Hamiltonian Eq.~(1), two numerical techniques were here applied.
One is the relaxation technique, in which the optimal positions of the 
oxygens are determined by minimizing the total energy.
In this calculation, only the stretching mode for the octahedron, 
namely $u_{\bf i}^{\bf a}+u_{\bf i-a}^{\bf a}=0$, is taken into account.
Moreover, the relaxation has been performed for fixed structures of 
the t$_{\rm 2g}$-spins such as ferro (F), A-type AF (A-AF), 
C-type AF (C-AF), and G-type AF (G-AF), shown in Fig.~1(a).
The advantage of this method is that the optimal orbital structure
can be rapidly obtained on small clusters.
However, the assumptions involved in the relaxation procedure
should be checked with an independent method.
Such a check is performed with the unbiased MC simulations
used before by our group.\cite{Yunoki2}
The dominant magnetic and orbital structures
are deduced from correlation functions.
In the MC method, the clusters currently reachable are
$2 \times 2 \times 2$, $4 \times 4 \times 2$, and 
$4\times 4 \times 4$.  In spite of this size limitation,
arising from the large number of degrees of 
freedom in the problem, the available clusters are sufficient for our
mostly qualitative purposes.
In addition, the remarkable agreement between MC and relaxation methods 
lead us to believe that our results 
are representative of the bulk limit.

\begin{figure}[h]
\vskip1.0truein
\hskip-1.7truein
\centerline{\epsfxsize=1.8truein \epsfbox{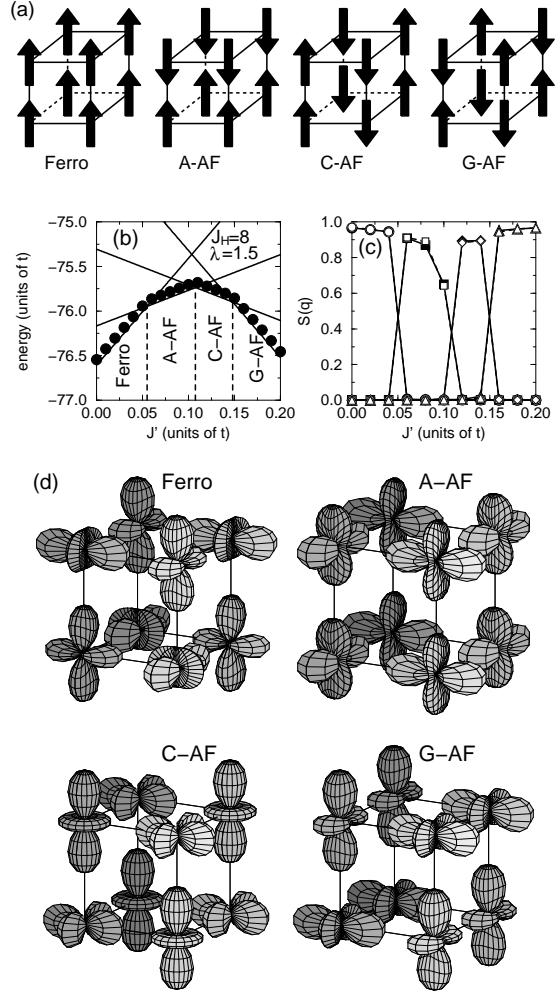} }
\vskip-1.2truein
\caption{
(a) Magnetic structures in $2\times 2 \times 2$ clusters.
(b) Total energy as a function of $J'$ on a
$2\times 2 \times 2$ lattice with $J_{\rm H}=8$ and $\lambda=1.5$.
The solid lines and circles indicate the relaxation and MC results,
respectively. MC simulations have been performed at temperature $1/200$.
(c) Spin correlation function $S({\bf q})$ obtained by
MC simulations as a function of $J'$, at $J_{\rm H}=8$ and $\lambda=1.5$.
Solid and open symbols denote the results
in $4\times 4 \times 2$ and $4\times 4 \times 4$ clusters, respectively.
Circles, squares, diamonds, and triangles indicates $S({\bf q})$ for
${\bf q}=(0,0,0)$, $(\pi,0,0)$, $(\pi,\pi,0)$,
and $(\pi,\pi,\pi)$, respectively.
(d) Optimized orbital structure for each magnetic structure.}
\label{fig1}
\end{figure}

In Fig.~1(b), the mean-energy is presented as a function of $J'$
for $J_{\rm H}=8$ and $\lambda=1.5$, on a $2 \times 2 \times 2$ cluster 
with open boundary conditions.
The solid lines and symbols indicate the results obtained with the 
relaxation technique and MC simulations, respectively.
The agreement is excellent, showing that the
relaxation method is accurate. The 
small deviations between the results of the two techniques are caused
by temperature effects.
As intuitively expected, with increasing $J'$ the optimal magnetic 
structure changes from ferro- to antiferromagnetic, and this
occurs in the order 
F$\rightarrow$A-AF$\rightarrow$C-AF$\rightarrow$G-AF.
To check size effects, the t$_{\rm 2g}$-spin
correlation function $S({\bf q})$ was calculated also in 
$4 \times 4 \times 2$ and $4\times 4 \times 4$ clusters,
where $S({\bf q})=
(1/N)\sum_{\bf i,j}e^{-i{\bf q}\cdot({\bf i}-{\bf j})}
\langle {\bf S}_{\bf i}\cdot{\bf S}_{\bf j} \rangle$,
$N$ is the number of sites, and $\langle \cdots \rangle$
indicates the thermal average value.
As shown in Fig.~1(c), with increasing $J'$ the dominant correlation 
changes in the order of ${\bf q}=(0,0,0)$, $(\pi,0,0)$, $(\pi,\pi,0)$, 
and $(\pi,\pi,\pi)$.
The values of $J'$ at which the spin structures changes
agree well with those in Fig.~1(b).

The shape of the occupied orbital arrangement with the lowest energy 
for each magnetic structure is in Fig.~1(d).
For the F-case, the G-type orbital structure is naively expected,
but actually a more complicated orbital structure is stabilized,
indicating the importance of the cooperative treatment 
for JT-phonons.
For the A-AF state,\cite{comment3}
only the C-type structure is depicted, but the 
G-type structure, obtained by a $\pi$/2-rotation of the upper $x$-$y$ plane
of the C-type state, was found to have {\it exactly} the same energy.
Small corrections will remove this degeneracy in favor of the C-type as
described below.
For C- and G-AF, the obtained orbital structures are
G- and C-types, respectively.
Although the same change of the magnetic structure due to $J'$ was 
already reported in the electronic model with purely Coulomb 
interactions,\cite{Maezono} 
the orbital structures in those previous
calculations were G-, G-, A-, and A-type for the
F-, A-AF, C-AF, and G-AF spin states, respectively.
Note that for the A-AF state, of relevance for the undoped manganites,
the G-type order was obtained,\cite{Maezono} 
although in another treatment for the Coulomb interaction,
the C- and G-type structures were found to be degenerate,\cite{Mizokawa}
as in our calculation. 

\begin{figure}[b]
\hskip-0.2truein
\vskip0.5truein
\centerline{\epsfxsize=2.5truein \epsfbox{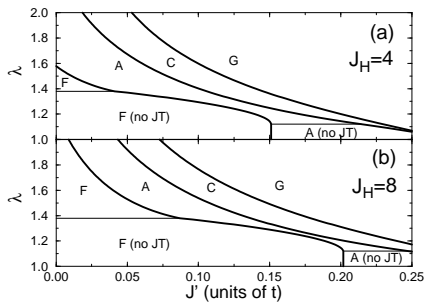} }
\vskip-1.truein
\label{fig2}
\caption{Magnetic phase diagram on the $(J',\lambda)$ plane
for (a) $J_{\rm H}=4$ and (b) $8$ (relaxation method).
Below the thin solid lines in the F and A-AF regions,
the JT-distortion disappears, suggesting that the system becomes metallic.
}
\end{figure}

In Figs.~2 (a) and (b), the phase diagrams 
on the $(J', \lambda)$-plane are shown 
for $J_{\rm H}=4$ and $8$, respectively.
The curves are drawn by the relaxation method.
As expected, the F-region becomes wider with increasing $J_{\rm H}$.
When $\lambda$ is increased at fixed $J_{\rm H}$, 
the magnetic structure changes from 
F$\rightarrow$A-AF$\rightarrow$C-AF$\rightarrow$G-AF.
This tendency is qualitatively understood if the two-site problem is 
considered in the limit $J_{\rm H} \gg 1$ and $E_{\rm JT} \gg 1$,
where $E_{\rm JT}$ is the static JT-energy given by
$E_{\rm JT}=\lambda^2/2$.
The energy-gain due to the second-order hopping process of 
e$_{\rm g}$-electrons 
is roughly $\delta E_{\rm AF} \sim 1/J_{\rm H}$ and 
$\delta E_{\rm F} \sim 1/E_{\rm JT}$ for AF- and F-spin pairs, respectively.
Increasing $E_{\rm JT}$, $\delta E_{\rm F}$ decreases,
indicating the relative stabilization of the AF-phase.
In our phase diagram, the A-AF phase appears for 
$\lambda \agt 1.1$ and $J'\alt 0.15$.
This region does not depend much on $J_{\rm H}$, as long as 
$J_{\rm H} \gg 1$.
Although $\lambda$ seems to be large, it is 
realistic from an experimental viewpoint:
$E_{\rm JT}$ is $0.25$eV from photoemission experiments\cite{Shen}
and $t$ is estimated as $0.2 \sim 0.5$eV,\cite{Saito}
leading to $1 \alt \lambda \alt 1.6$.
As for $J'$, it is estimated as $0.02 \alt J' \alt 0.1$.\cite{Ishihara,Perring}
Thus, the location in
parameter-space of the 
A-AF state found here
is reasonable when compared with experimental results for LaMnO$_3$.

\begin{figure}[h]
\vskip0.3truein
\hskip-0.2truein
\centerline{\epsfxsize=3.0truein \epsfbox{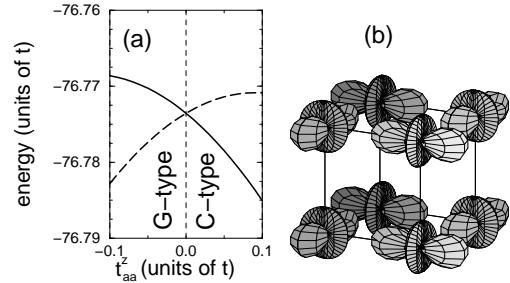} }
\vskip-1.2truein
\label{fig3}
\caption{(a) Total energy as a function of $t_{\rm aa}^{\bf z}$
on the $2\times 2 \times 2$ lattice with $L_{\bf x}=L_{\bf y}>L_{\bf z}$
for $J_{\rm H}=8$, $\lambda=1.6$
and $J'=0.05$ (relaxation method).
The solid and dashed curves denote the C- and
G-orbital states, respectively.
(b) Orbital structure in the A-AF phase
for $t_{\rm aa}^{\bf z}=0^{+}$ and $L_{\bf x}=L_{\bf y}>L_{\bf z}$.
These shapes are very close to purely $3x^2-r^2$ and $3y^2-r^2$ types.}
\end{figure}

Let us now focus on the orbital structure in the A-AF phase.
In the cubic lattice studied thus far, the 
C- and G-type orbital structures are degenerate, and it is unclear
whether the orbital pattern in the $x$-$y$ plane corresponds
to the alternation of $3x^2-r^2$ and $3y^2-r^2$ orbitals 
observed in experiments.\cite{LaMnO3}
To remedy the situation, some empirical facts observed 
in manganites become important:
(i) The MnO$_6$ octahedra are slightly tilted from 
each other, leading to modifications in the hopping matrix.
Among these modifications, the generation of a
 non-zero value for $t_{\rm aa}^{\bf z}$ 
is important.
(ii) The lattice is not cubic, but the relation 
$L_{\bf x} \approx L_{\bf y} > L_{\bf z}$ holds. From 
experimental results,\cite{LaMnO3} these numbers are estimated
as $L_{\bf x}=L_{\bf y}=4.12$\AA~and $L_{\bf z}=3.92$\AA,
indicating that the distortion with $Q_3$-symmetry occurs spontaneously.
Note that the hopping amplitude and $J'$ along the $z$-axis
are different from those in the $x$-$y$ plane due to this distortion.
Motivated by these observations, the energies for C- and G-type 
orbital structures were recalculated including this time a nonzero value
for $t_{\rm aa}^{\bf z}$ in the magnetic A-AF state (see Fig.~3(a)).
In the real material, it can be shown based on symmetry considerations
that the tilting of the MnO$_{6}$ 
octahedra will always lead to a {\it positive} value for $t_{\rm aa}^{\bf z}$.
Then, the results of Fig.~3(a) 
suggest that the C-type orbital structure should 
be stabilized in the real materials, and the
 explicit shape of the occupied orbitals is shown in Fig.~3(b).
The experimentally relevant 
C-type structure with the approximate alternation of $3x^2-r^2$ and 
$3y^2-r^2$ orbitals is indeed successfully obtained by this procedure.
Although the octahedron tilting actually leads to a change of all hopping
amplitudes, effect not including in this work,
the present analysis is sufficient to show that
the C-type orbital structure is stabilized in the A-AF magnetic phase when 
$t_{\rm aa}^{\bf z}$ is a small positive number,\cite{comment2} as it occurs
in the real materials.

In this work the Coulomb interactions (intra-orbital Coulomb $U$, 
inter-orbital Coulomb $U'$, and inter-orbital exchange $J$)
have been neglected.
For Mn-oxides, they are estimated as $U=7$eV, $J=2$eV, and 
$U'=5$eV,\cite{Ishihara} which are large compared to $t$.
However, the result for the optimized distortion described in this
paper, obtained without the Coulomb interactions, is not expected 
to change since the energy gain due to the JT-distortion is maximized 
when a single e$_{\rm g}$-electron is present per site. This is 
essentially the same effect as produced by a short-range repulsion.
In fact, it has been checked explicitly 
by using the Exact Diagonalization method 
that the JT- and breathing-distortions are not changed by $U'$ on a 
$2 \times 2$ cluster using the F-state in which $U$ and $J$ 
can be neglected.
In addition, the MC simulations show that the probability of doble
occupancy of a single orbital is negligible where the
A-type spin, C-type orbital state is stable.\cite{Benedetti}
Based on all these observations, it is believed that that the effect of 
the Coulomb interaction is not crucial for the appearance of the A-AF
state with the proper orbital order. Another way to rationalize this
result is that the integration of the JT-phonons at large
$\lambda$ will likely induce Coulombic interactions dynamically.

Finally, let us briefly discuss transitions induced
by the application of external magnetic fields on undoped manganites.
When the A-AF state is stabilized, 
the energy difference (per site) obtained in our study
between the A-AF and F states is about $t/100$.
As a consequence, magnetic fields of $20 \sim 50$T 
could drive the transition from A-AF to F order
accompanied by a change of orbital structure, 
interesting effect which 
may be observed in present magnetic field facilities.

Summarizing, using numerical techniques at $n=1$ 
it has been shown that 
the A-AF state is stable in models with
JT-phonons, using coupling values
physically reasonable for LaMnO$_3$. Our results indicate
that it is not necessary
to include large Coulombic interactions in the calculations to 
reproduce experimental results for the manganites.
Considering the small but important effect of the
 octahedra tilting of the real materials, 
the C-type orbital structure (with the alternation pattern of
$3x^2-r^2$ and $3y^2-r^2$ orbitals) has been successfully reproduced
for the A-AF phase in this context.

T.H. thanks Y. Takada and H. Koizumi for enlightening discussion.
T.H. is supported from the Ministry of Education, Science,
Sports, and Culture of Japan. 
E.D. is supported by grant NSF-DMR-9814350.


\end{multicols}

\begin{references}
\bibitem{comment} In this work, the buckling and rotational modes 
of MnO$_6$ octahedron are not explicitly included.
\bibitem{comment2} It can be shown that this mechanism to stabilize the 
C-type structure works also for the purely electronic model in the 
Hartree-Fock approximation.
\end{references}
\end{document}